# A Stochastically Evolving Non-local Search and Solutions to Inverse Problems with Sparse Data

Mamatha Venugopal[1], Ram Mohan Vasu[1], and Debasish Roy[2*]

[1]Optical Tomography Lab, Department of Instrumentation and Applied Physics, Indian Institute of Science, Bangalore – 560012, India

[2]Computational Mechanics Lab, Department of Civil Engineering, Indian Institute of Science, Bangalore – 560012, India

*Corresponding Author, email: royd@civil.iisc.ernet.in

## Abstract

*Building upon our earlier work of a martingale approach to global optimization, a powerful stochastic search scheme for the global optimum of cost functions is proposed on the basis of change of measures on the states that evolve as diffusion processes and splitting of the state-space along the lines of a Bayesian game. To begin with, the efficacy of the optimizer, when contrasted with one of the most efficient existing schemes, is assessed against a family of $N_p$-hard benchmark problems. Then, using both simulated- and experimental data, potentialities of the new proposal are further explored in the context of an inverse problem of significance in medical imaging, wherein the superior reconstruction features of a global search vis-à-vis the commonly adopted local or quasi-local schemes are brought into relief.*

**Keywords:** *global optimization, inverse problems, martingales, Bayesian games*

## 1. Introduction

Inverse problems aim at the recovery of unknown parameters of a system, typically a mathematical model given perhaps by a set of differential equations, based on a few noisy measurements of the system response. Solutions to inverse problems may yield crucial parameter information with potential applications in many areas of science and engineering. Despite the exciting possibilities, a generally agreed numerical framework enabling acceptable solutions to inverse problems remains elusive, partly owing to so called non-uniqueness, as it arises in a deterministic setting (a regularized quasi-Newton method to wit) engendered by model and data (measurement) insufficiency. Moreover, presence of noise in the data may cause such solutions to drift to infeasible regions. A basic recipe for solving an inverse problem is the minimization of an objective functional that specifies the misfit between the available measurements and the predictions from the recovered model. In a deterministic setup involving sufficiently smooth fields, a common way to perform this minimization is through a gradient-based local search as exemplified, say, by the iterative Gauss-Newton (GN) method [1]. A GN-based scheme necessarily incorporates certain regularization strategies [2] that impose *a priori* constraints on the inverse problem to yield stable and meaningful solutions. Here the choice of 'right' regularization parameters adds to the computational burden brought in by Jacobian calculations in nonlinear problems. Indeed most objective functions, being non-convex, multimodal and perhaps non-differentiable, preclude the very applicability of a GN-like scheme. In contrast, a Bayesian search scheme [3] founded on the probability theory affords a more natural means to account for the numerous possible solutions by allowing the underlying probability distribution to be multimodal. Starting with an assumed prior, the aim of such a scheme would be to estimate the posterior parameter distribution conditioned on the noisy measurements. An approach that incorporates Bayesian updates is based on the filtered martingale problem [4-5] wherein the parameter to be recovered is treated as a stochastic process [6], possibly with respect to an iteration variable in case the system is time-independent. It has been shown that this approach enables obtaining additive updates to the parameters based on a change of measures so as to drive the resulting measurement-prediction misfit to a zero-mean martingale [6-7]. Convergence to a martingale structure ensures that the expectation of the measurement-prediction misfit, treated as a stochastic process, will remain zero and invariant to random perturbations during subsequent iterations or temporal recursions. It is known that, under fairly general conditions, the solution to a filtered martingale problem is unique [8] and this is perhaps a welcome departure from the non-uniqueness issues that confront a deterministic setup. Nevertheless, given that solutions could be highly sensitive to data noise, model errors and varying dimensions of the data and parameter sets, a filtered martingale problem, numerically implemented through a Monte Carlo scheme involving a finite ensemble, may at best ensure that the objective function attains an available local minimum. Moreover, upon averaging over a multi-modal posterior distribution where some of the modes may not even be physically relevant, the recovered estimates for the system parameters could be in significant error. This problem is exacerbated with sparse data availability, a case often encountered in practice. A more rational strategy could be in the form of a stochastically founded non-local or global search scheme to pick out the most relevant mode in the posterior distribution or, perhaps to redefine a modified distribution around this mode and thus address the deterioration of the quality of solutions owing to averaging over multiple possibilities.

Numerous heuristic and meta-heuristic global optimization schemes [9] abound the literature, prominently including genetic algorithm (GA) [10], simulated annealing [11], particle swarm optimization (PSO) [12], differential evolution (DE) [13] and covariance matrix adaptation evolution strategy (CMA-ES) [14] to name a few. Most such evolutionary schemes begin with a random scatter of candidate solutions, henceforth referred to as particles, which evolve over subsequent iterations according to a scheme-specific update strategy. The update steps aim at enabling the particles to explore the state space in order to detect the global minimum of the objective function. Success depends on the right *exploration-exploitation* trade-off that charts out a middle path between computationally expensive exploration and quickly identifying the global extremum from amongst the available extrema. While a full exposition of various schemes is not within our scope, brief outlines of a few prominent ones should be in order. The GA and CMA-ES assign weights to each particle based on the 'closeness' of the computed objective function to its available optimal value. In particular, only the best fit particles spawn new ones at a subsequent iteration. Such a weight-based approach that neglects the 'bad' particles might lead to a faster yet premature convergence to a local extremum despite the exploratory steps involved. This problem, known as 'particle collapse' in the stochastic filtering parlance, occurs when the entire weight is assumed by a single particle as the iterations progress. This is clearly demonstrated for the case of CMA-ES while attempting to minimize some of the benchmark objective functions in Section 5. In partial amelioration of this bottleneck, schemes like DE and PSO apply heuristically derived additive corrections to particles in the update stage.

Interestingly, none of the schemes discussed so far are naturally equipped to handle multivariate/multi-objective optimization (MOO), the *sine qua non* in solving many inverse problems. Although there have been attempts at adapting CMA-ES, one of the most powerful evolutionary schemes, for MOO, only limited success in applications has accrued [15]. In [16-17], we proposed a generalized optimization framework, COMBEO (Change Of Measure Based Evolutionary Optimization), based mainly on a perturbed martingale problem that could rationally accommodate updates by different existing methods within a single mathematical structure. The need for such a unified framework was inspired by the no free lunch theorems [18] that proved the near impossibility of a single optimization scheme performing well across the spectrum of $N_P$-hard problems. Yet another advantage of

COMBEO was in its inherent ability to treat multi-objective problems. On the downside, COMEO either required a large ensemble size or an inflated number of measurements to solve a given problem. One of our current aims is thus to modify COMBEO so as to better equip it to solve practical problems with smaller ensemble sizes and sparser sets of measurements. This is primarily accomplished by incorporating within the martingale problem of local optimization, a new update strategy based on state space splitting (3S). Additionally, perturbative exploratory steps such as scrambling, blending etc. guide the greedy local search to converge to the global optimum. Our second focus here is on establishing, using experimental data, the efficacy of a non-local search as encoded within COMEO for parameter reconstruction and contrast such performance with that of a more localized search as represented, say, by COMBEO stripped off its global search tools. In the process, we demonstrate the usefulness of the 3S scheme in solving inverse problems with sparse measurements.

The rest of the paper is organized as follows. Section 2 poses optimization as a filtered martingale problem and puts forth the bare-bones additive update strategy that renders the measurement-prediction misfit a zero-mean martingale. The random exploratory operations that aid in the global search are briefly explained in Section 3. Section 4 contains a game theoretic perspective that leads to the 3S scheme followed by a pseudo-code for the proposed evolutionary search. The first part of Section 5 gives comparative results of the proposed scheme vis-à-vis CMA-ES in minimizing a few benchmark objective functionals. Thereafter, we undertake a numerical study to contrast the present approach vis-à-vis one without the trappings of global search in the context of a medical imaging problem, viz. quantitative photoacoustic tomography. The aim is to bring out, perhaps for the first time, what a well-conceived global search scheme can do when the available measurements are very sparse. Reconstructions for both simulated and experimental data are given. The concluding remarks are presented in Section 6.

## 2. Optimization as a filtered martingale problem

For a nonlinear objective functional $f(\mathbf{x}): \mathbb{R}^{n_x} \to \mathbb{R}$, our aim is to find $\mathbf{x}^{\min} \in \mathbb{R}^{n_x}$ such that $f(\mathbf{x}^{\min}) \leq f(\mathbf{x}) \forall \mathbf{x} \in \mathbb{R}^{n_x}$. For the multi-objective case with $\mathbf{f}(\mathbf{x}) := \left(f^1(\mathbf{x}), ..., f^{n_f}(\mathbf{x})\right) \in \mathbb{R}^{n_f}$, each component $f^i(\mathbf{x}) \forall i = 1, ..., n_f$ has to be minimized. If we wish to solve the deterministically posed optimization problem by borrowing ideas from stochastic filtering, the parameters to be recovered and the objective functions must be treated as stochastic processes, which are possibly of the diffusion type evolving with respect to a time-like iteration variable $\tau$. Within a complete probability space $(\Omega, \mathcal{F}, \mathcal{F}_\tau, P)$ [6], such a characterization would render $\mathbf{x}: \Omega \to \mathbb{R}^{n_x}$ an $n_x$-dimensional random vector at each iteration with $\mathcal{F}$ denoting the Borel σ-algebra over open subsets of $\mathbb{R}^{n_x}$, $\mathcal{F}_\tau$ the natural filtration and P the probability measure. Noting however that the parameters are usually not governed by any dynamics, $\mathbf{x}$ may have to be artificially evolved as a continuous Brownian motion or discrete random walk in $\tau$ (evolution with jump discontinuities, modelled as Levy processes, is possible and even desirable for more efficient non-local search; but not considered here). The evolution of the continuously parameterized stochastic process $\mathbf{x}_\tau$ is then represented in the form of a stochastic differential equation (SDE) [6]:

$$d\mathbf{x}_\tau = d\mathbf{B}_\tau \qquad (2.1)$$

Here, $\mathbf{B}_\tau \in \mathbb{R}^{n_x}$ is a vector Brownian motion with mean zero and covariance matrix $\Sigma_B \Sigma_B^T \in \mathbb{R}^{n_x \times n_x}$. Although $\tau$ is by definition monotonically increasing in $\mathbb{R}^+$, in practice, $\mathbf{x}_\tau$ is evolved only over finite increments of $\tau$, i.e. for $\tau_1 < ... < \tau_{n_t}$. Thus, for the $(k+1)^{th}$ iteration where $\tau \in (\tau_k, \tau_{k+1}]$, we can write a discrete form of equation (2.1) as

$$\mathbf{x}_{k+1} = \mathbf{x}_k + \Delta \mathbf{B}_k \qquad (2.2)$$

where $\Delta \mathbf{B}_k = \mathbf{B}_{k+1} - \mathbf{B}_k$. Let a minimum of the objective function be denoted as $\tilde{\mathbf{f}} = \mathbf{f}(\mathbf{x}^{min})$. In a deterministic setup, when the design variables approach their extremal value, the first variation of the objective function tends to zero. In a non-local set-up based on stochastic updates, we may intuitively extend this notion to require that, conditioned on $\tilde{\mathbf{f}}$, $\mathbf{f}(\mathbf{x}_\tau)$ is mean-invariant during subsequent iterations, i.e. the error $\tilde{\mathbf{f}} - \mathbf{f}(\mathbf{x}_\tau)$ becomes a zero-mean martingale, e.g. an Ito integral, as $\mathbf{x}_\tau$ approaches $\mathbf{x}^{min}$. Keeping consistency with a stochastic filtering scheme [7,19], one may then write a measurement-like equation to pose the problem of optimization as

$$\tilde{\mathbf{f}}_\tau = \mathbf{f}(\mathbf{x}_\tau) + \mathbf{W}_\tau \qquad (2.3)$$

Though $\mathbf{W}_\tau \in \mathbb{R}^{n_f}$ should ideally be an Ito integral, keeping in mind the equivalence of an Ito integral with a Brownian motion through a proper scaling of $\tau$ that we do not explicitly effect here, $\mathbf{W}_\tau$ is considered a vector Brownian motion itself with mean zero and covariance matrix $\Sigma_W \Sigma_W^T \in \mathbb{R}^{n_f \times n_f}$. Also, in most practical cases, the extremal value of $\mathbf{f}$ is *a-priori* unknown and in such a scenario, $\tilde{\mathbf{f}}_\tau$ denotes the available minimum from a finite ensemble of computed function values, upon substitution of the particles. Having formulated the filtering problem, the aim is to estimate $\pi_\tau(\mathbf{x}) := E[\mathbf{x}_\tau | \mathcal{G}_\tau]$ where $E[.]$ denotes the expectation operator with respect to the underlying measure $P$ and $\mathcal{G}_\tau$ the filtration generated by $\tilde{\mathbf{f}}_\tau$ whose distribution is likely to be multi-modal.

Denoting by $n_e$ the ensemble size in the Monte Carlo (MC) setup that we adopt here, the initial set of particles $\{\hat{\mathbf{x}}_0(j)\}_{j=1}^{n_e}$ is generated from a random scatter in the state space. Thereafter, the updated set of particles, $\{\hat{\mathbf{x}}_k(j)\}_{j=1}^{n_e}$ at any iteration $k = 0,..., MAX-1$ follows a prediction-update strategy to obtain the solution at $k+1$.

*Prediction:* The particle-wise prediction equation follows from equation (2.2):

$$\mathbf{x}_{k+1}(j) = \mathbf{x}_k(j) + \Delta \mathbf{B}_k(j), j = 1,...,n_e \qquad (2.4)$$

*Update:* The update step involves estimation of the filtered conditional distribution $\pi_\tau(\mathbf{x})$, which solves the Kushner-Stratonovich (KS) equation [20] in nonlinear filtering. An essential ingredient *en route* to solving the KS equation is an SDE form of the measurement. It can be shown that the KS-based update equation reduces to that based on a Kalman filter [21] when the measurements are linear and Gaussian. In order to avoid the aberrant approximations involved in casting equation (2.3) as an SDE, in the present work, we follow a simpler route that emulates Kalman-like updates. Specifically, we employ an update strategy consistent with an ensemble square root filter [22], which could be considered as a nonlinear version of the Kalman filter. Each predicted particle is iteratively updated through an additive gain-like correction term that drives the error $\tilde{\mathbf{f}}_\tau - \mathbf{f}(\mathbf{x}_\tau)$, also called innovation in the filtering parlance, to a zero-mean martingale. The update equation, written particle-wise, is given as

$$\hat{\mathbf{x}}_{k+1}(j) = \mathbf{x}_{k+1}(j) + \mathbf{G}_{k+1}\left(\tilde{\mathbf{f}}_{k+1} - \mathbf{f}(\mathbf{x}_{k+1}(j))\right), j = 1,\ldots,n_e \quad (2.5)$$

$\mathbf{G}_{k+1} \in \mathbb{R}^{n_x \times n_f}$ is the gain matrix computed as $\mathbf{G}_{k+1} = \mathbf{X}_{k+1}\mathbf{F}_{k+1}^T\left(\mathbf{F}_{k+1}\mathbf{F}_{k+1}^T + \Sigma_W \Sigma_W^T\right)^{-1}$ where $\mathbf{X}_{k+1} \in \mathbb{R}^{n_x \times n_e}$ and $\mathbf{F}_{k+1} \in \mathbb{R}^{n_f \times n_e}$ are ensemble perturbation matrices respectively given by

$$\mathbf{X}_{k+1} = \frac{1}{\sqrt{n_e - 1}}\left[\mathbf{x}_{k+1}(1) - \frac{1}{n_e}\sum_{j=1}^{n_e}\mathbf{x}_{k+1}(j),\ldots,\mathbf{x}_{k+1}(n_e) - \frac{1}{n_e}\sum_{j=1}^{n_e}\mathbf{x}_{k+1}(j)\right]$$

and

$$\mathbf{F}_{k+1} = \frac{1}{\sqrt{n_e - 1}}\left[\mathbf{f}(\mathbf{x}_{k+1}(1)) - \frac{1}{n_e}\sum_{j=1}^{n_e}\mathbf{f}(\mathbf{x}_{k+1}(j)),\ldots,\mathbf{f}(\mathbf{x}_{k+1}(n_e)) - \frac{1}{n_e}\sum_{j=1}^{n_e}\mathbf{f}(\mathbf{x}_{k+1}(j))\right].$$

Also, $\tilde{\mathbf{f}}_{k+1}$ and $\mathbf{f}(\mathbf{x}_{k+1}(j))$ denote respectively the available function minimum within the ensemble and the computed function value of the $j^{th}$ particle at the $(k+1)^{th}$ iteration. The parameter estimate is empirically approximated as:

$$\overline{\mathbf{x}}_{k+1}(n_e) = \frac{1}{n_e}\sum_{j=1}^{n_e}\hat{\mathbf{x}}_{k+1}(j) \quad (2.6)$$

Note that the update equation (2.5) is similar in form to a GN-based update [3], wherein the employed Fréchet derivative bears functional analogy with the gain matrix. Thus one could interpret $\mathbf{G}_{k+1}$ as directing a derivative-free, non-local search, a crucial aspect that is missing in most evolutionary schemes. The uniqueness of the update stems from its roots in the powerful machinery of stochastic calculus. Moreover, this approach to optimization naturally accommodates multi-objective functions. However despite its inherent non-locality, the stochastic search described above is ill-equipped to track down all the peaks in the posterior distribution, unless the ensemble size is large enough. Increased noise intensity in the prediction step (equation (2.4)) could have helped combing the parameter space better, were it not for the unacceptably wild scatter of the predicted particles based on Brownian increment, which has an unbounded variation. Here again, one may try to constrain the noise increments by using Doob's *h*-transform [23]; but the computational overhead would be prohibitive. Thus the update in equation (2.5), with low noise intensity in equation (2.4), should be viewed as a quasi-local search. In the next couple of sections, therefore, we consider a few global search tools that could be employed effectively and efficaciously.

## 3. Random Perturbation schemes

Our random exploratory steps, which provide additional layers of random search beyond that enabled by the quasi-local scheme considered in Section 2, aim at tracking the global extremum by avoiding local traps. Thus, at any iteration, a particle generated via equation (2.5) is subject to a perturbation scheme that is imposed through a sequence of inner iterations indexed by $i$ such that when the perturbations vanish we arrive at the updated particle, i.e. $\mathbf{x}_{k+1}(j) = {}^0\mathbf{x}_{k+1}(j) \to {}^1\mathbf{x}_{k+1}(j) \to ... \to {}^i\mathbf{x}_{k+1}(j) \to ... \to {}^\infty\mathbf{x}_{k+1}(j) = \hat{\mathbf{x}}_{k+1}(j)$. In practice, we set an upper bound for $i$, i.e. $i < i_{max}$. Generally, the iterative update procedure indexed by $k$ coincides with the inner iterations and hence the left superscript $i$ may be removed from the notations. The perturbation schemes considered in this section are *coalescence*, *scrambling*, *blending* and *relaxation*, each of which is briefly described in the following subsections.

### 3.1 Coalescence

Solution to a global optimization problem requires that the posterior probability distribution associated with the converged solution be unimodal irrespective of the objective function profile. Ideally, then, the converged measure $\pi_\tau(.)$ should correspond to a Dirac measure $\delta_{\mathbf{x}_{min}}(.)$ where $\mathbf{x}_{min}$ denotes the global optimum. The unimodality constraint is generally not satisfied by a quasi-local scheme, such as stochastic filtering, and the resulting multimodal profile of the recovered density (if it exists) has a peak corresponding to each detected local extremum. The unimodality constraint on the probability density could be imposed by coalescence, i.e. by forcing the particles to evolve according to a Wiener martingale whose mean is given by the available global extremum. Such a characterization could be straightaway incorporated within the martingale problem as an additional innovation by allowing the noisy scatter to evolve as a zero-mean martingale upon convergence.

Let the noisy scatter be represented by $\left\{ \mathbf{x}_{k+1}(j) - \mathbf{x}_{k+1}(\sigma_1(j)) \right\}_{j=1}^{n_e}$ where $\sigma_1(j)$ denotes a random index from the set $\{1,...,n_e\} \setminus \{j\}$. In order to implement coalescence, consider the update of the $j^{th}$ particle. We wish to drive the error $\mathbf{x}_{k+1}(j) - \mathbf{x}_{k+1}(\sigma_1(j))$ to, say, a zero-mean Brownian motion, $\mathbf{W}_{k+1}^c$ with a very low intensity. This is accomplished by inflating the innovation vector in equation (2.5) as $\mathbf{I}_{k+1}(j) = \begin{bmatrix} \tilde{\mathbf{f}}_{k+1} - \mathbf{f}(\mathbf{x}_{k+1}(j)) \\ \mathbf{x}_{k+1}(j) - \mathbf{x}_{k+1}(\sigma_1(j)) \end{bmatrix} \in \mathbb{R}^{(n_f + n_x) \times 1}$. The cost filtration $\mathcal{G}_\tau$ is suitably modified to incorporate the sub-filtration generated by $\mathbf{W}^c$. The new update equation then becomes

$$\hat{\mathbf{x}}_{k+1}(j) = \mathbf{x}_{k+1}(j) + \mathbf{G}_{k+1} \mathbf{I}_{k+1}(j), j = 1, ..., n_e \tag{3.1}$$

With a convenient notational abuse, we retain here the same notation for $\mathbf{G}_{k+1}$ although its expression now becomes $\mathbf{G}_{k+1} = \mathbf{X}_{k+1} \mathbf{F} \mathbf{X}_{k+1}^T (\mathbf{F} \mathbf{X}_{k+1} \mathbf{F} \mathbf{X}_{k+1}^T + Cov_{meas})^{-1}$ where

$$\mathbf{FX}_{k+1} = \frac{1}{\sqrt{n_e - 1}} \left[ \mathbf{I}_{k+1}(1) - \frac{1}{n_e} \sum_{j=1}^{n_e} \mathbf{I}_{k+1}(j), ..., \mathbf{I}_{k+1}(n_e) - \frac{1}{n_e} \sum_{j=1}^{n_e} \mathbf{I}_{k+1}(j) \right]$$

and $Cov_{meas} = \begin{bmatrix} \Sigma_W \Sigma_W^T & 0 \\ 0 & \alpha \mathcal{I}(n_x) \end{bmatrix}$. $\alpha \in \mathbb{R}^+$, $\alpha \ll 1$ corresponds to the intensity of $\mathbf{W}_{k+1}^c$ and $\mathcal{I}(n_x)$ denotes an identity matrix of dimension $n_x \times n_x$.

The ease with which coalescence could be included in the update equation signifies one of the prominent advantages of posing optimization as a martingale problem. The non-unique choice of the innovation vector unfolds possibilities of improving upon the current scheme and thus designing powerful global search schemes.

### 3.2 Scrambling

The coalescence alone is inadequate in pulling out the particles stuck in local traps, a possibility rendered likely owing to the very small updates imparted to particles around the extremal (local or global) values. A stalling of the search scheme may be circumvented by scrambling, i.e. by randomly swapping the updates of two particles. In this work, we implement scrambling as follows:

$$\hat{\mathbf{x}}_{k+1}(j) = \mathbf{x}_{k+1}(\sigma_2(j)) + \mathbf{G}_{k+1} \mathbf{I}_{k+1}(j), j = 1, ..., n_e \tag{3.2}$$

where $\sigma_2(j)$ denotes a random index from the set $\{1, ..., n_e\} \setminus \{j\}$. Furthermore, borrowing from a basic idea of DE [13], one may employ element-wise scrambling that executes the swapping operation separately for each of the $n_x$ components of the vector $\mathbf{x}_{k+1}(j)$, thereby allowing the particles to assume a larger spectrum of variations. The resulting update equation may then be written as

$$\hat{x}_{k+1}^{l}(j) = x_{k+1}^{l}\left(\sigma_{2}^{l}(j)\right) + U_{k+1}^{l}(j), \quad l = 1,...,n_x, j = 1,...,n_e \tag{3.3}$$

where $\mathbf{x}_{k+1} = \left(x_{k+1}^{1},...,x_{k+1}^{n_x}\right)$ etc. and $U_{k+1}^{l}(j)$ is the $l^{th}$ component of the update vector $\mathbf{U}_{k+1}(j) := \mathbf{G}_{k+1}\mathbf{I}_{k+1}(j)$. The superscript in $\sigma_{2}^{l}(j)$ indicates that for every particle, each of the $n_x$ components receive updates from the corresponding components of a different particle.

As we have observed, while coalescence imposes the unimodality constraint, scrambling tries to avoid the possible stalling of the search scheme. The mutually competing goals of the two strategies might result either in a premature collapse of the solutions to a single point or a drift of solutions to infeasible regions. The coalescence-scrambling dilemma is reflective of the need for yet another layer of randomness that would simultaneously ensure that the particles are repelled from the local valleys and, once at the global extremum, do not drift apart.

### 3.3 Blending

The idea for blending originated from the 'crossover' technique in genetic algorithms [10] wherein an offspring is created by the fusion of two individual particles. Even so, a blended particle at any iteration is generated by a linear combination of the original particle and its update rather than as a fusion of two original particles, where the term 'original' refers to the predicted particles at the given iteration.

$$\hat{\mathbf{x}}_{k+1}^{b}(j) = w_{k+1}(j)\mathbf{x}_{k+1}(j) + \left(1 - w_{k+1}(j)\right)\hat{\mathbf{x}}_{k+1}(j), \quad j = 1,...n_e \tag{3.4}$$

The weights $w_{k+1}(.)$ are calculated based on a suitably defined fitness function $\chi_{k+1}(.)$, given below, that quantifies the measurement-prediction misfit.

$$\chi_{k+1}(j) = \sqrt{\left(\tilde{f}_{k+1}^{1} - f^{1}\left(\mathbf{x}_{k+1}(j)\right)\right)^{2} + ... + \left(\tilde{f}_{k+1}^{n_f} - f^{n_f}\left(\mathbf{x}_{k+1}(j)\right)\right)^{2}} \tag{3.5}$$

Also, define $\tilde{w}_{k+1}(j) = \sum_{m=1}^{n_e} \chi_{k+1}(m) w_k(m) - \chi_{k+1}(j) w_k(j)$. The weights $w_{k+1}(.)$ are obtained by normalizing $\tilde{w}_{k+1}(.)$ to sum to 1. Equation (3.4) may be rearranged to obtain

$$\left|\hat{\mathbf{x}}_{k+1}^b(j) - \mathbf{x}_{k+1}(j)\right| = (1 - w_{k+1}(j))\left|\hat{\mathbf{x}}_{k+1}(j) - \mathbf{x}_{k+1}(j)\right|$$

i.e. $\left|\hat{\mathbf{x}}_{k+1}^b(j) - \mathbf{x}_{k+1}(j)\right| \leq \left|\hat{\mathbf{x}}_{k+1}(j) - \mathbf{x}_{k+1}(j)\right|$

The above inequality and the imposition of the coalescence step render $\left|\hat{\mathbf{x}}_{k+1}^b(j) - \mathbf{x}_{k+1}(j)\right|$ a converging sequence, i.e., given $\delta > 0, \exists K \in \mathbb{N}$ such that for $k > K$, $\left|\hat{\mathbf{x}}_{k+1}^b(j) - \mathbf{x}_{k+1}(j)\right| < \delta$ almost surely. One may thus draw the following conclusions.

(a) At any iteration $k$, let the $l^{th}$ and $m^{th}$ particles have the highest and lowest values of $\chi$; then $w_{k+1}(l) \leq w_{k+1}(m)$ and $P\left(\hat{\mathbf{x}}_{k+1}^b(m) = \mathbf{x}_{k+1}(m)\right) \geq P\left(\hat{\mathbf{x}}_{k+1}^b(l) = \mathbf{x}_{k+1}(l)\right)$.

(b) Upon convergence, blending chooses the original particle with higher probabilities.

Further implications of blending are detailed in the next section from a game theoretic perspective.

### 3.4 Relaxation and Selection

In the presence of diffusive noises, inclusion of random exploratory steps will almost surely introduce 'bad' particles that correspond to infeasible solutions. This might necessitate the introduction of a few selection criteria to discard the badly behaved particles. Even with scrambling and blending, a search scheme that always chooses to update the particles would quickly collapse to a local trap owing to the highly directed and coalesced search imposed by the martingale problems. Such a scenario is prevented by the so-called relaxation, i.e. by assigning positive probabilities to the event of retaining particles without updates. Denoting the inertia factor by $p_I$, this ensures that the particles are updated with a probability $(1 - p_I) < 1$. The probability of regular and blended updates is each $\frac{(1 - p_I)}{2}$. Additionally, the selection step chooses only those updates that reduce $\chi_{k+1}(j)$ and if the update increases the fitness value, the original particle is retained. Since the selection step performs the role of functional minimization indirectly, in some global optimization problems, one may eliminate the error terms $\tilde{\mathbf{f}}_{k+1} - \mathbf{f}(\mathbf{x}_{k+1}(j))$ from the innovation vector. In such cases, the prediction step might also become redundant as the particles are allowed ample exploration by the perturbation schemes such as scrambling and blending.

## 4. State space splitting and a game-theoretic interpretation of optimization

### 4.1 State Space Splitting (3S)

The dimensionality curse, which besets many stochastic search schemes including most stochastic filters with the necessity of an exponentially exploding ensemble size, comes in the way of solving inverse problems with a large number of unknowns. To a large extent, the particle degeneracy problems encountered by weight-based schemes are circumvented in the additive-update strategies that attempt at 'healing' the bad particles instead of eliminating them altogether [7]. However, even with such an approach, the quality of solutions is highly dependent on the ensemble size, $n_e$, as has been proved in [24]. The ensemble finiteness limits the search space which could possibly lead to premature convergence of the solution to a local minimum. Hence, an increase in the system dimension $n_x$ has to be inevitably accompanied with an increase in $n_e$ to effect a proper exploration of the state space. This increase is likely orders of magnitude higher as opposed to a linear increase owing to the slow convergence rate, $\dfrac{1}{\sqrt{n_e}}$ [25], of the MC simulation. A possible amelioration of particle explosion would be by splitting the original problem into smaller parts and solve for each lower dimensional component separately. A divide and solve approach is justified on a fairly accurate assumption that a larger ensemble is required to solve the original problem compared to the locally-split problem to achieve the same level of accuracy. The 3S scheme may be incorporated on the following lines.

a) Set the number of sub-problems/substructures, $n_p$; each substructure corresponding to an $\dfrac{n_x}{n_p}$-dimensional state vector. If $n_x$ is not divisible by $n_p$, then the first $n_p - 1$ substructures will have $floor\left(\dfrac{n_x}{n_p}\right)$ components and the remaining state variables constitute the last substructure. Denote by $\mathbf{s}^{(1)},...,\mathbf{s}^{(n_p)}$ the $n_p$ substructures and by $n_s^{(m)}$ the size of the $m^{th}$ substructure.

b) Following element-wise scrambling, each substructure is updated sequentially as follows:

$$\hat{s}_{k+1}^{(m),l}(j) = s_{k+1}^{(m),l}\left(\sigma_2^l(j)\right) + U_{k+1}^{(m),l}(j), \quad l=1,...,n_x, j=1,...,n_e, m=1,...,n_p \quad (4.1)$$

where $\hat{s}_{k+1}^{(m),l}(j)$ and $s_{k+1}^{(m),l}(j)$ denote respectively the updated and predicted components of the $j^{th}$ particle of the $l^{th}$ state that is in the $m^{th}$ substructure. Also, $\mathbf{U}_{k+1}^{(m)}(j) = \mathbf{G}_{k+1}^{(m)} \mathbf{I}_{k+1}^{(m)}(j)$ is the update vector corresponding to the $m^{th}$ substructure and is computed by assimilating the last update information of the other $(m-1)$ substructures. Specifically, $\mathbf{G}_{k+1}^{(m)} \in \mathbb{R}^{n_s^{(m)} \times (n_f + n_x)}$ is given by

$$\mathbf{G}_{k+1}^{(m)} = \mathbf{S}_{k+1}^{(m)} \left(\mathbf{FX}_{k+1}^{(m)}\right)^T \left(\mathbf{FX}_{k+1}^{(m)} \left(\mathbf{FX}_{k+1}^{(m)}\right)^T + Cov_{meas}\right)^{-1}$$

where $\mathbf{S}_{k+1}^{(m)} = \dfrac{1}{\sqrt{n_e - 1}} \left[ \mathbf{s}_{k+1}^{(m)}(1) - \dfrac{1}{n_e} \sum_{j=1}^{n_e} \mathbf{s}_{k+1}^{(m)}(j), ..., \mathbf{s}_{k+1}^{(m)}(n_e) - \dfrac{1}{n_e} \sum_{j=1}^{n_e} \mathbf{s}_{k+1}^{(m)}(j) \right]$

and $\mathbf{FX}_{k+1}^{(m)} = \dfrac{1}{\sqrt{n_e - 1}} \left[ \mathbf{I}_{k+1}^{(m)}(1) - \dfrac{1}{n_e} \sum_{j=1}^{n_e} \mathbf{I}_{k+1}^{(m)}(j), ..., \mathbf{I}_{k+1}^{(m)}(n_e) - \dfrac{1}{n_e} \sum_{j=1}^{n_e} \mathbf{I}_{k+1}^{(m)}(j) \right]$.

Also, $\mathbf{I}_{k+1}^{(m)}(j) = \begin{bmatrix} \tilde{\mathbf{f}}_{k+1} - \mathbf{f}\left(\mathbf{x}_{k+1}^{(m)}(j)\right) \\ \mathbf{x}_{k+1}^{(m)}(j) - \mathbf{x}_{k+1}^{(m)}\left(\sigma_1(j)\right) \end{bmatrix}$

and $\mathbf{x}_{k+1}^{(m)}(j) = \begin{bmatrix} \hat{\mathbf{s}}_{k+1}^{(1)}(j) & \cdots & \hat{\mathbf{s}}_{k+1}^{(m-1)}(j) & \mathbf{s}_{k+1}^{(m)}(j) & \cdots & \mathbf{s}_{k+1}^{(n_p)}(j) \end{bmatrix}^T$.

c) The blended update could be similarly written as

$$\hat{\mathbf{s}}_{k+1}^{(m),b}(j) = w_{k+1}^{(m)}(j)\mathbf{s}_{k+1}^{(m)}(j) + \left(1 - w_{k+1}^{(m)}(j)\right)\hat{\mathbf{s}}_{k+1}^{(m)}(j), \quad j=1,...n_e, m=1,...,n_p \quad (4.2)$$

where the weights $w_{k+1}^{(m)}(.)$ are obtained by normalizing $\tilde{w}_{k+1}^{(m)}(.)$ to sum to 1 and

$\tilde{w}_{k+1}^{(m)}(j) = \sum_{l=1}^{n_e} \chi_{k+1}^{(l)}(l) w_{k+1}^{(l-1)}(l) - \chi_{k+1}^{(m)}(j) w_{k+1}^{(m-1)}(j), m > 1$ and $w_{k+1}^{1}(j) = w_{k+1}(j)$ where $w_{k+1}(j)$

are the weights after prediction at the $(k+1)^{th}$ step. Moreover, the fitness value is calculated as $\chi_{k+1}^{(m)}(j) = \sqrt{\left(\tilde{f}_{k+1}^1 - f^1\left(\mathbf{x}_{k+1}^{(m)}(j)\right)\right)^2 + \ldots + \left(\tilde{f}_{k+1}^{n_f} - f^{n_f}\left(\mathbf{x}_{k+1}^{(m)}(j)\right)\right)^2}$.

*Remarks*:

i. Although the update, regular or blended, itself is done separately, the correlation between the substructures is maintained by incorporating the last available information from the rest within the update of one.

ii. Evident from the expression for $\mathbf{I}_{k+1}^{(m)}$ is the fact that the innovation vector for the $m^{th}$ substructure involves the entire set of state variables rather than the $n_s^{(m)}$ components belonging to it. This could be contrasted with the localization techniques used in [24].

iii. The finally updated particle is obtained by concatenating the individual updates, i.e.
$$\hat{\mathbf{x}}_{k+1}(j) = \left[\hat{\mathbf{s}}_{k+1}^{(1)}(j), \ldots, \hat{\mathbf{s}}_{k+1}^{(n_p)}(j)\right]^T.$$

The current implementation of the 3S requires $\mathcal{O}(n_p \times n_e)$ functional evaluations at each recursion of the algorithm as opposed to $\mathcal{O}(n_e)$ evaluations that would have sufficed if one were to solve for $\mathbf{x}$ without splitting. Nevertheless, the extra functional evaluations are more than justified given that the ensemble size $n_E$ required to solve the original problem would almost always be far higher than $n_p \times n_e$. This is corroborated by the numerical examples wherein some of the functional minimization problems of dimension 40 could not be solved by CMA-ES even with ensemble sizes as high as 1000 whereas the proposed scheme with $n_p = 2, 4$ etc. solved them with $n_e = 20$. Another notable advantage of the 3S is that it allows for a more exhaustive assimilation of data owing to the inner iterations in the split-updates. This feature is especially handy while solving higher-dimensional inverse problems with sparse data. Using inner iterations in the original problem would have required a far larger computational overhead. Note, however, that the performance of the 3S scheme deteriorates as the number of substructures is increased beyond a threshold for a given dimension. This could be owing to the inadequate data communication between the substructures as they grow in number. Thus, for the best results, a balance must be struck between the number of substructures and the ensemble size.

## 4.2 Game-theoretic interpretations

The optimization scheme could be set forth as a Bayesian game which would be insightful and put the various steps in a better perspective. A non-cooperative game [26] consists of two or more interacting decision makers or players, a set of strategies or actions for each player and cost functions over the sets of action profiles. If each player is aware of the optimal (cost-minimizing) strategies of all other players, the game is one with perfect information. In contrast, games with incomplete information wherein at least one of the players does not know the preferences of at least another are modelled as Bayesian games [26]. Comparing our global optimization scheme with a non-cooperative game necessarily requires a game-theoretic equivalent of a global minimum.

For an $n$-player non-cooperative game, a set of strategies and their corresponding costs constitute a Nash Equilibrium (NE) [27] if no player can reduce cost by deviating from the strategy forming the NE. The current optimization scheme, fashioned after a multi-player Bayesian game, need to identify the global minimum with a Bayesian NE. Let the $n_p$ substructures be the players, forming the index set $\mathcal{N} = \{1,...,n_p\}$. The appropriateness of Bayesian modelling arises from the unavailability, *a-priori*, of the global optimum, so the players are unsure of their optimal strategies at any given time. A Bayesian game identifies each player as a parameter subset, $\mathbf{s}^{(m)} : \Omega \to \mathbb{R}^{n_s^{(m)}}$, $m=1,...,n_p$, so that $\left[\mathbf{s}^{(1)},...,\mathbf{s}^{(n_p)}\right]^T = \mathbf{x}$. This is followed by the *a-priori* assignment of a probability measure $P$. The ideal representation of these random variables would be via an infinite ensemble $\left\{\mathbf{s}^{(m)}(j)\right\}_{j=1}^{\infty}$, $m=1,...,n_p$ and perhaps a uniform measure to begin with, which are updated as the game progresses depending on the weight $p\left(\tilde{\mathbf{f}}_\tau \mid \mathbf{s}_\tau^{(m)}(j)\right)$. However, in practice, the ensemble is only finite, say of size $n_e$ and the goal is to choose strategies so as to drive the particles to the global optimum. Specifically, a player can choose from a set of pure and mixed strategies as follows.

a) *Pure strategies:*
   i. $\mathrm{a}^o$ : The player retains her original position without updates.
   ii. $\mathrm{a}^u$ : The player updates her coordinates according to equation (4.1).

b) *Mixed strategies:*

$a^b$: The player chooses a blended update according to equation (4.2).

A randomizer assigns probabilities $p_I$, $\dfrac{(1-p_I)}{2}$ and $\dfrac{(1-p_I)}{2}$ respectively for the strategies $a^o$, $a^u$ and $a^b$. The inertia factor $p_I$ prevents the player from moving to a local equilibrium (LE) point (a local minimum) although at the cost of slower convergence. The strategy $a^u$ uses coalescence and scrambling for the update, inducing the player to minimize her cost based on the game history until that point. Note that if the players were to choose $a^u$ with high probability, then the game might end prematurely at an LE. Unlike the NE in a standard game, no player can do better in the present setup by deviating from an LE owing to a highly unimodal posterior distribution imposed by coalescence. To a large extent, the stalling at an LE could be prevented by the randomizer. The blended update, a weighted mixture of pure strategies, increases the particle diversity and hence the information entropy of the underlying distribution. Denoting by $p_\tau^{(m),o}$ and $p_\tau^{(m),u}$ the probabilities assigned to $a^o$ and $a^u$ respectively by the $m^{th}$ player at $\tau$, we have $p_\tau^{(m),o} = w_\tau^{(m)}$ and $p_\tau^{(m),u} = 1 - w_\tau^{(m)}$ satisfying $p_\tau^{(m),o} + p_\tau^{(m),u} = 1 \ \forall m$. Also, denote an $n_p$-tuple of strategies by $\boldsymbol{a}_\tau = \{a_\tau^1,...,a_\tau^{n_p}\}$ where $a^i \in \{a^o, a^u, a^b\}, i = 1,...,n_p$. Let $\mathcal{A}^{n_p}$ denote the set of all possible $n_p$-tuples. If one works with a cost function (as in stochastic control), a tolerance limit $\varepsilon < 1$ may be set that determines the level of accuracy to which a given problem needs to be solved. The game ends when the tolerance is weakly reached or when each player has played $MAX$ turns. The cost function $C: \mathcal{A}^{n_p} \times \Omega \to \mathbb{R}$ is a random variable that determines the preferences of the players to the possible strategies. The cost of the $m^{th}$ player is given by $C_\tau^{(m)} := \chi_\tau^{(m)}$ which is bounded below by $0$. Unlike the convention used with stochastic control, here the cost, by definition, is an implicit function of the strategies. For instance, if $\boldsymbol{a}_\tau = \{a^o,...,a^o\}$, then $\chi_\tau^{(m)} = \chi_{\tau-}^{(m)}, m = 1,...,n_p$ and so on. To be consistent with the game theoretic parlance, one could formally write $C_\tau^{(m)} := c_\tau\left(a^{1*},...,a^{(m-1)*},a^m,...,a^{n_p*}\right)$, where the superscript * indicates that the $m^{th}$ player makes her move assuming that all others play their optimal strategies.

It is insightful to note that the strategies may be identified as (possibly implicit) Borel functions $\mathbf{g}(\mathbf{x}_\tau, \mathbf{a}_\tau)$. Assuming $\|\mathbf{g}(\mathbf{x}_\tau, \mathbf{a}_\tau)\| \leq N \in \mathbb{N} \, \forall \tau \in [0, \tau_{MAX}]$, one can use Doob's $h$-transform [28] to define an equivalent change of measure $\mathrm{P} \to \mathrm{P}^a$ using the Radon-Nikodym derivative $\Lambda_\tau$ given by

$$\Lambda_\tau = \mathrm{E}\left[\frac{d\mathrm{P}^a}{d\mathrm{P}}\bigg|\mathcal{F}_\tau\right] := \exp\left\{\int_0^\tau \mathbf{g}(\mathbf{x}_s, \mathbf{a}_s).d\mathbf{B}_s - \frac{1}{2}\int_0^\tau |\mathbf{g}(\mathbf{x}_s, \mathbf{a}_s)|^2 \, ds\right\}$$

This in turn renders $\mathbf{B}_\tau^a = \mathbf{B}_\tau - \int_0^\tau \mathbf{g}(\mathbf{x}_s, \mathbf{a}_s) ds$ a $\mathrm{P}^a$-Brownian motion adapted to $\mathcal{F}_\tau$ and the original state SDE acquires a drift to become $d\mathbf{x}_\tau = \mathbf{g}(\mathbf{x}_\tau, \mathbf{a}_\tau) d\tau + d\mathbf{B}_\tau^a$, the solution to which may be numerically obtained using one of the available schemes [34]. This characterization does not need a stochastic projection as in the filtering theory and therefore marks a departure from the optimization setup in Section 2. This could be a particularly useful interpretation, even though an explicit form for $\mathbf{g}$ might not be available for all strategies.

For an inverse problem, the cost function typically encapsulates a normed discrepancy of the computed solution and the data. The NE for the game then corresponds to the case when $\tilde{\mathbf{f}}_\tau - \mathbf{f}(\mathbf{x}_\tau)$ is a zero-mean martingale. Equivalently, consistent with [29] and along the lines of stochastic control, the notion of optimality may also be brought out by defining an integrated cost, i.e. $C_\tau^{(m)} = \int_0^\tau \chi_s^{(m)}(\boldsymbol{a}) ds + \inf_{\boldsymbol{a} \in \mathcal{A}^{n_p}} \int_0^{\tau^* \wedge \tau_{MAX}} \chi_s^{(m)}(\boldsymbol{a}) ds$ wherein the strategy set $\boldsymbol{a}$ could be viewed as a control variable. Here, $\tau^*$ denotes a stopping time corresponding to $\mathrm{E}_\mathrm{P}\left[C_\tau^{(m)}\right] < \varepsilon, m = 1,...,n_p$. Clearly, $C_\tau^{(m)}$ is a $\mathrm{P}^a$-submartingale by definition; it becomes a $\mathrm{P}^a$-martingale only when optimal strategies are employed, i.e. when $\boldsymbol{a} = \boldsymbol{a}^*$ (though not presented here, a proof for this would be on similar line as in [29]).

Finally, we note that the current work deals only with a classical form of non-local search that is consistent with Bell's inequalities. States with a quantum or no-signalling type nonlocality, modelled using the noncommutative probability theory, might endow the present optimization setup with newer notions of the global optimum.

We end this section with a pseudo-code of the proposed algorithm.

### 4.3 Pseudo-code

Set $n_e, n_p$ and $MAX$. If $n_p$ is divisible by $n_x$, then $n_s^{(m)} = \dfrac{n_x}{n_p}$, else each of the $n_p - 1$ substructures will have $floor\left(\dfrac{n_x}{n_p}\right)$ components and the remaining $n_x - (n_p - 1) n_s^{(1)}$ unknowns form the last substructure. Set the inertia factor $p_I$. Initialize the blending coefficients $w_0(j) = \dfrac{1}{n_e}, j = 1,...,n_e$. Generate the initial ensemble of particles, $\{\hat{\mathbf{x}}_0(j)\}_{j=1}^{n_e}$ based on an initial, possibly uniform, distribution in the admissible search space. The search space could be $\mathbb{R}^{n_x}$ or a bounded space, $[lb, ub]^{n_x}$ where $lb$ and $ub$ denote the lower and upper bounds respectively. Also calculate the initial fitness values, $\chi_0(j), j = 1,...,n_e$. Set $k = 0$.

1. *(Prediction)* Obtain the predicted set of particles $\{\mathbf{x}_{k+1}(j)\}_{j=1}^{n_e}$ using equation (2.4).

2. Generate $\boldsymbol{\sigma}_1 = (\sigma_1(1),...,\sigma_1(n_e))$ and $\boldsymbol{\sigma}_2 = (\sigma_2(1),...,\sigma_2(n_e))$. Set $m = 1$.

3. Calculate $\mathbf{f}\left(\mathbf{x}_{k+1}^{(m)}(j)\right)$, $\chi_{k+1}^{(m)}(j)$ and $w_{k+1}^{(m)}(j)$ for $j = 1,...,n_e$. Note that for $m = 1$,

    $\mathbf{f}\left(\mathbf{x}_{k+1}^{(m)}(j)\right) = \mathbf{f}\left(\mathbf{x}_{k+1}(j)\right)$, $\chi_{k+1}^{(m)}(j) = \chi_{k+1}(j)$ and $w_{k+1}^{(m)}(j) = w_{k+1}(j)$. Set $j = 1$.

4. If $rand < (1 - p_I)$, then

    if $rand < 0.5$

    update $s_{k+1}^{(m),l}(j)$ according to equation (4.1) to obtain $\hat{s}_{k+1}^{(m),l}(j)$ for $l = 1,...,n_x$.

    Else

    update $\mathbf{s}_{k+1}^{(m)}(j)$ according to equation (4.2) to obtain $\hat{\mathbf{s}}_{k+1}^{(m),b}(j)$. Put $\hat{\mathbf{s}}_{k+1}^{(m)}(j) = \hat{\mathbf{s}}_{k+1}^{(m),b}(j)$.

    Else, retain the original particle, i.e. $\hat{\mathbf{s}}_{k+1}^{(m)}(j) = \mathbf{s}_{k+1}^{(m)}(j)$.

5. Set $j = j + 1$. If $j \leq n_e$, go to step 4; else go to step 6.

6. Construct $\mathbf{x}_{k+1}^{(m+1)}(j) = \begin{bmatrix} \hat{\mathbf{s}}_{k+1}^{(1)}(j) & \cdots & \hat{\mathbf{s}}_{k+1}^{(m)}(j) & \mathbf{s}_{k+1}^{(m+1)}(j) & \cdots & \mathbf{s}_{k+1}^{(n_p)}(j) \end{bmatrix}^T$. Set $m = m+1$.

   If $m \leq n_p$, go to step 3; else go to step 7.

7. Construct $\hat{\mathbf{x}}_{k+1}(j) = \begin{bmatrix} \hat{\mathbf{s}}_{k+1}^{(1)}(j), \ldots, \hat{\mathbf{s}}_{k+1}^{(n_p)}(j) \end{bmatrix}^T$ for $j = 1, \ldots, n_e$ and calculate the empirical mean according to equation (2.6).

8. Set $k = k+1$. If $k < MAX$, go to step 1; else terminate the algorithm.

## 5. Numerical Illustrations

### 5.1 Benchmark Problems

Now we aim at assessing the performance of the proposed scheme against a few benchmark minimization problems. The level of complexity in solving a given problem typically increases with increase in system dimension as the multimodal nature of a cost function varies considerably with the number of design variables. For instance, the Rastrigin function has roughly $10^{n_x}$ local optima [30]. For a given dimension, the difficulty levels may vary across problems owing to specific characteristics such as the degree of separability [30] of the cost function. We consider two sets of benchmark functions in our relative assessment exercise involving the proposed scheme vis-à-vis CMA-ES, one of the most successful global optimization schemes till date. In all the cases, the number of design variables, $n_x = 40$, the ensemble size, $n_e = 20$, the inertia factor, $p_I = 0.9$ and the tolerance, $\varepsilon = 10^{-5}$. The first set consisting of 20 benchmark functions (F1-F20) is the test suite that was released for the CEC'2010 special session and competition on large-scale global optimization [30]. The various functions are as follows:

a) Separable functions: shifted elliptic function (F1), shifted Rastrigin function (F2), shifted Ackley function (F3)

b) Single-group $n$-non-separable functions: single-group shifted and $n$-rotated elliptic function (F4), single-group shifted and $n$-rotated Rastrigin function (F5), single-group shifted and $n$-rotated Ackley function (F6), single-group shifted $n$-dimensional Schwefel's problem 1.2 (F7), single-group shifted $n$-dimensional Rosenbrock function (F8)

c) $\frac{n_x}{2n}$-group $n$-nonseparable functions: $\frac{n_x}{2n}$-group shifted and $n$-rotated elliptic function (F9), $\frac{n_x}{2n}$-group shifted and $n$-rotated Rastrigin function (F10), $\frac{n_x}{2n}$-group shifted and $n$-rotated Ackley function (F11), $\frac{n_x}{2n}$-group shifted $n$-dimensional Schwefel's problem 1.2 (F12), $\frac{n_x}{2n}$-group shifted $n$-dimensional Rosenbrock function (F13)

d) $\frac{n_x}{n}$-group $n$-nonseparable functions: $\frac{n_x}{n}$-group shifted and $n$-rotated elliptic function (F14), $\frac{n_x}{n}$-group shifted and $n$-rotated Rastrigin function (F15), $\frac{n_x}{n}$-group shifted and $n$-rotated Ackley function (F16), $\frac{n_x}{n}$-group shifted $n$-dimensional Schwefel's problem 1.2 (F17), $\frac{n_x}{n}$-group shifted $n$-dimensional Rosenbrock function (F18)

e) Nonseparable functions: shifted Schwefel's problem 1.2 (F19), shifted Rosenbrock function (F20)

The mathematical formulae and specific characteristics of these functions maybe found in [30] wherein $D$ and $m$ must be read as $n_x$ and $n$ respectively. The proposed scheme takes four substructures ($n_p = 4$) to minimize these functions, all of which have their global minimum at 0. Table 1 gives the comparative performance of the proposed scheme and CMA-ES in minimizing F1-F20. It is noted that except for F14, the current proposal is able to locate the global minimum to the desired level of accuracy whereas CMA-ES fails in as many as 7 cases. Out of the 7 failed cases, CMA-ES could solve F2, F10, F15 etc with $n_e \sim 1000$ whereas it failed to minimize some other functions such as F6 even with still higher ensemble sizes. Worthwhile is also the fact that the proposed scheme with $n_e = 20, n_p = 1$ could not minimize F6 as the algorithm always got trapped in the local well corresponding to $F6 \in [19, 20]$, indicating the usefulness of the 3S strategy in solving higher dimensional optimization problems with smaller ensembles sizes.

The second set of 24 functions (IF1-IF24) have been taken from the document on black-box optimization benchmarking [31]. The functions are categorized as

a) Separable functions: sphere function (IF1), ellipsoidal function (IF2), Rastrigin function (IF3), Büche-Rastrigin function (IF4), linear slope (IF5)

b) Functions with low/moderate conditioning: attractive sector function (IF6), step ellipsoidal function (IF7), original Rosenbrock function (IF8), rotated Rosenbrock function (IF9)

c) Functions with high conditioning and unimodal: ellipsoidal function (IF10), discus function (IF11), bent cigar function (IF12), sharp ridge function (IF13), different powers function (IF14)

d) Multimodal functions with adequate global structure: Rastrigin function (IF15), Weierstrass function (IF16), Schaffers F7 function (IF17), Schaffers F7 function, moderately ill-conditioned (IF18), composite Griewank-Rosenbrock function F8F2 (IF19)

e) Multimodal functions with weak global structure: Schwefel function (IF20), Gallagher's Gaussian 101-me peaks function (IF21), Gallagher's Gaussian 21-hi peaks function (IF22), Katsuura function (IF23), Lunacek bi-Rastrigin function (IF24)

Here, the maximum number of iterations is set to $4 \times 10^5$ and $n_p = 2$. Table 2 reports the comparative results of the proposed scheme and CMA-ES in minimizing functions IF1-IF24. The results indicate the higher difficulty level of these problems as several of them could not be solved by either of the methods. Nevertheless, except for functions IF7 and IF12, the proposed scheme scores over CMA-ES either in solving the problem completely or in arriving at lesser errors. The inability of CMA-ES to converge to the global optimum is in many cases owing to its greedy weight-based search that directs the particles to one of the local traps. Hence, although we put *MAX* in cases where CMA-ES has failed, the search might have stalled, having reached the reported error much earlier in the process. This may be contrasted with the proposed scheme wherein the error gradually decreases in most cases as the iterations progress, revealing the possibility that it might reach the optimal value asymptotically. The greedy nature of CMA-ES also accounts for its faster convergence in solving functions such as F8, F9, IF8 etc. This fact has been numerically verified during our assessment exercises with the benchmark functions although a detailed exposition is out of the present scope. At this point, the results in Tables 1 and 2 are also worth comparing with those via the original version of COMBEO [17]. The results indicate substantive improvement in the search scheme, manifest both in the level of accuracy and in the substantively reduced ensemble sizes.

Table 1. Performance of the proposed scheme and CMA-ES against objective functions F1-F20; $n_x = 40, n_e = 20, \varepsilon = 10^{-5}, MAX = 4 \times 10^5$.

| Objective Function | Proposed Scheme | | CMA-ES | |
|---|---|---|---|---|
| | Number of Iterations | Error | Number of Iterations | Error |
| F1 | 709 | $\varepsilon$ | 3581 | $\varepsilon$ |
| F2 | 5438 | $\varepsilon$ | MAX | 64.26 |
| F3 | 721 | $\varepsilon$ | 572 | $\varepsilon$ |
| F4 | 4505 | $\varepsilon$ | 5390 | $\varepsilon$ |
| F5 | 6419 | $\varepsilon$ | MAX | 9.95 x $10^5$ |
| F6 | 921 | $\varepsilon$ | MAX | 1.03 x $10^7$ |
| F7 | 614 | $\varepsilon$ | 2739 | $\varepsilon$ |
| F8 | 21387 | $\varepsilon$ | 4065 | $\varepsilon$ |
| F9 | 218671 | $\varepsilon$ | 4792 | $\varepsilon$ |
| F10 | 46073 | $\varepsilon$ | MAX | 39.8 |
| F11 | 212709 | $\varepsilon$ | MAX | 19.8 |
| F12 | 855 | $\varepsilon$ | 388 | $\varepsilon$ |
| F13 | 76824 | $\varepsilon$ | 9976 | $\varepsilon$ |
| F14 | MAX | 934.6 | 5614 | $\varepsilon$ |
| F15 | 145270 | $\varepsilon$ | MAX | 12.9 |
| F16 | 3426 | $\varepsilon$ | MAX | 15.2 |
| F17 | 1301 | $\varepsilon$ | 436 | $\varepsilon$ |
| F18 | MAX | 0.0058 | 28005 | $\varepsilon$ |
| F19 | 15050 | $\varepsilon$ | 1033 | $\varepsilon$ |
| F20 | 22476 | $\varepsilon$ | 4566 | $\varepsilon$ |

Table 2. Performance of the proposed scheme and CMA-ES against INRIA objective functions IF1-IF20; $n_e = 20, \varepsilon = 10^{-5}, MAX = 1.6 \times 10^5$.

| Objective Function | Target Function Value | Proposed Scheme | | CMA-ES | |
|---|---|---|---|---|---|
| | | Number of Iterations | Error | Number of Iterations | Error |
| IF1 | 79.48 | 384 | $\varepsilon$ | 216 | $\varepsilon$ |
| IF2 | -209.88 | 572 | $\varepsilon$ | 3611 | $\varepsilon$ |
| IF3 | -462.09 | 4955 | $\varepsilon$ | MAX | 6.28 x $10^1$ |
| IF4 | -462.09 | 6517 | $\varepsilon$ | MAX | 9.9 x $10^1$ |
| IF5 | -9.21 | 84 | $\varepsilon$ | 976 | $\varepsilon$ |
| IF6 | 35.9 | 4282 | $\varepsilon$ | 1198 | $\varepsilon$ |
| IF7 | 92.94 | MAX | 2.89 | MAX | 1.58 x $10^1$ |
| IF8 | 149.15 | 15623 | $\varepsilon$ | 4300 | $\varepsilon$ |
| IF9 | 123.83 | MAX | 3.33 x $10^{-2}$ | 4026 | $\varepsilon$ |
| IF10 | -54.94 | MAX | 1.19 x $10^2$ | 3761 | $\varepsilon$ |
| IF11 | 76.27 | 91480 | $\varepsilon$ | 2115 | $\varepsilon$ |
| IF12 | -621.11 | MAX | 8.07 | 1712 | $\varepsilon$ |

| IF13 | 29.97 | *MAX* | $2.7 \times 10^{-1}$ | *MAX* | $9.9 \times 10^{-1}$ |
| IF14 | -52.35 | *MAX* | $6.84 \times 10^{-5}$ | 1643 | $\varepsilon$ |
| IF15 | 1000 | *MAX* | $1.7 \times 10^{2}$ | *MAX* | $6.29 \times 10^{1}$ |
| IF16 | 71.35 | *MAX* | 1.95 | *MAX* | 2.48 |
| IF17 | -16.94 | *MAX* | $2.34 \times 10^{-3}$ | *MAX* | $1.58 \times 10^{-1}$ |
| IF18 | -16.94 | *MAX* | $4.13 \times 10^{-1}$ | *MAX* | $6 \times 10^{-1}$ |
| IF19 | -102.55 | *MAX* | 3.74 | *MAX* | 9.54 |
| IF20 | -546.5 | *MAX* | $3.15 \times 10^{-1}$ | *MAX* | 2.45 |
| IF21 | 40.78 | 21965 | $\varepsilon$ | *MAX* | $6.2 \times 10^{1}$ |
| IF22 | -1000 | *MAX* | 1.96 | *MAX* | 2.43 |
| IF23 | 6.87 | *MAX* | 1.31 | *MAX* | 5.67 |
| IF24 | 102.61 | *MAX* | $1.96 \times 10^{2}$ | *MAX* | $3.89 \times 10^{2}$ |

## 5.2 Quantitative Photoacoustic Tomography (QPAT): simulations and experiments

The goal of quantitative photoacoustic tomography (QPAT) is to recover the distribution of the optical absorption coefficient ($\mu_a(\mathbf{r})$) in soft tissues by measuring the photoacoustic pressure on the surface of a soft tissue-like object [32]. QPAT is very useful in the early detection of cancer as the absorption in the malignant cells could be 4-5 times higher than in the surrounding healthy tissue, thereby making imaging possible. The recovery of $\mu_a(\mathbf{r})$ from the measured boundary pressure is a two-step process involving the solution of two partial differential equations (PDEs), viz. the wave equation modelling the propagation of acoustic wave and the photon transport equation. Light pulses in the 700-1100 nm region used to illuminate an object, say **D**, causes absorption of light leading to an elevation in temperature and a consequent thermo-elastic expansion. Under conditions of thermal and stress confinements, the thermo-elastic expansion causes a localized pressure rise, $p_0(\mathbf{r}) = \Gamma \mu_a(\mathbf{r}) \phi(\mathbf{r}) = \Gamma H(\mathbf{r})$ which, in turn, sets off acoustic propagation in the object that is measured at the boundary using ultrasound transducers. Here, $\phi(\mathbf{r})$ denotes the light fluence, $\Gamma$ the Grüneisen coefficient and $H(\mathbf{r})$ the absorbed light energy density. The wave propagation equation is given by

$$\nabla^2 p(\mathbf{r},t) - \frac{1}{c^2}\frac{\partial^2 p(\mathbf{r},t)}{\partial t^2} = 0 \qquad (5.1)$$

Equation (5.1) is solved together with $\frac{\partial \rho_0(\mathbf{r})}{\partial t} = 0$. The wave equation serves as the forward model to recover $\rho_0(\mathbf{r})$ and consequently $H(\mathbf{r})$ from $\rho(\mathbf{r}), \mathbf{r} \in \partial \mathbf{D}$. A diffusive approximation to the photon transport equation relating $\mu_a(\mathbf{r})$ to $\phi(\mathbf{r})$ is given by

$$-\nabla \cdot \kappa \nabla \phi(\mathbf{r}) + \mu_a(\mathbf{r})\phi(\mathbf{r}) = S(\mathbf{r}) \qquad (5.2)$$

where $\kappa$ is the diffusion coefficient and $S(\mathbf{r})$ the incident light source term. See [32-33] for a detailed description of QPAT, the photoacoustic process and the underlying PDEs. The present aim is a direct recovery of $\mu_a(\mathbf{r}), \mathbf{r} \in \mathbf{D}$ from simulated and experimental $\rho(\mathbf{r}), \mathbf{r} \in \partial \mathbf{D}$. Specifically, we intend to study the implications of an efficient global search as contrasted with a quasi-local search, exemplified here by stochastic filtering.

The simulations are undertaken to study the effect the number of measurements has on the reconstruction accuracy. For both simulations and experiments, we consider a circular object of radius 12cm. The object in figure 1 is taken as the reference for the simulations where the anomalous region has $\mu_a = 0.05 \, \text{mm}^{-1}$ and the background $\mu_a^b = 0.01 \, \text{mm}^{-1}$. The object is uniformly illuminated from all sides with the strength of the source being 1. The detectors are placed uniformly around a circle of radius 2cm from the centre of the object and the measurements were collected for a total duration of $60 \, \mu$s at a sampling rate of 40 MHz. The values considered for the rest of the parameters are: $\kappa = 0.33 \, \text{mm}^{-1}, \Gamma = 1, c = 1500 \, \text{ms}^{-1}$. The simulated measurements are generated by adding 1% noise to the boundary pressure measurements. Reconstructions are performed using a triangular mesh with 313 nodes and by solving the forward equations in (5.1) and (5.2) by the finite element method. The ensemble size $n_e = 25$ is adopted for both filtering and optimization. Figures (2) and (3) show the reconstructions obtained respectively using filtering and optimization for the cases when the number of measurements is 51 and 25. While the filtering employs a prediction-update strategy given by equations (2.4)-(2.5), the optimization scheme follows the pseudo-code in Section 4.3 with $n_p = 2$ except for omitting coalescence from the innovation vector. Reconstructions by optimization are clearly superior in terms of quantitative accuracy and the recovered shape of the inclusion with fewer background artefacts. We note here that CMA-ES could not solve this problem even with higher ensemble sizes.

The experimental phantom had 3 inhomogeneous inclusions with $\mu_a^b = 0.01\,\text{mm}^{-1}$, $\kappa = 0.33$. The common $\mu_a$ value for two of the inhomogeneities is 0.04 mm$^{-1}$ and that for the third one is 0.05 mm$^{-1}$. There are 192 detectors placed uniformly around a circle of radius 4cm from the centre of the object and the measurements are collected over a total duration of 44.56 $\mu$s at a sampling rate of 50MHz. A more detailed account of the experimental setup and a schematic diagram may be found in [35]. Reconstructions are arrived at over a triangular mesh consisting of 1243 nodes using an ensemble of 50 particles. Given our current focus on the performance assessment of the proposed global optimizer with a parsimonious set of particles, the calibration details of the experiment are omitted. Figure 4 gives the reconstructions for the relative $\mu_a$ distribution normalized to [0,1] range. The central inhomogeneity that is missing in the filtering-based reconstruction is clearly visible in the reconstruction through optimization. While we can get a still better recovery by increasing $n_e$, figure 4(b) does offer an idea of the relative contrast and locations of the anomalies.

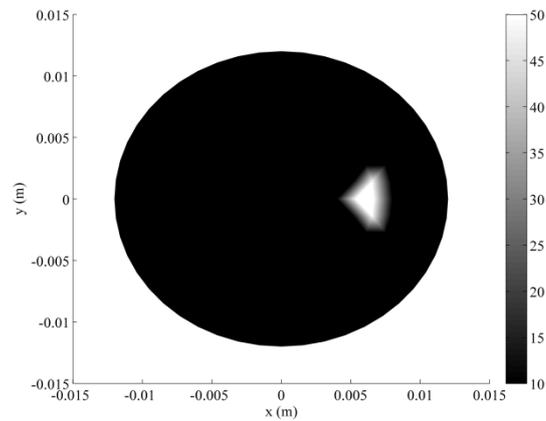

Figure 1. Distribution of $\mu_a$ (m$^{-1}$) in the simulated object (Reference)

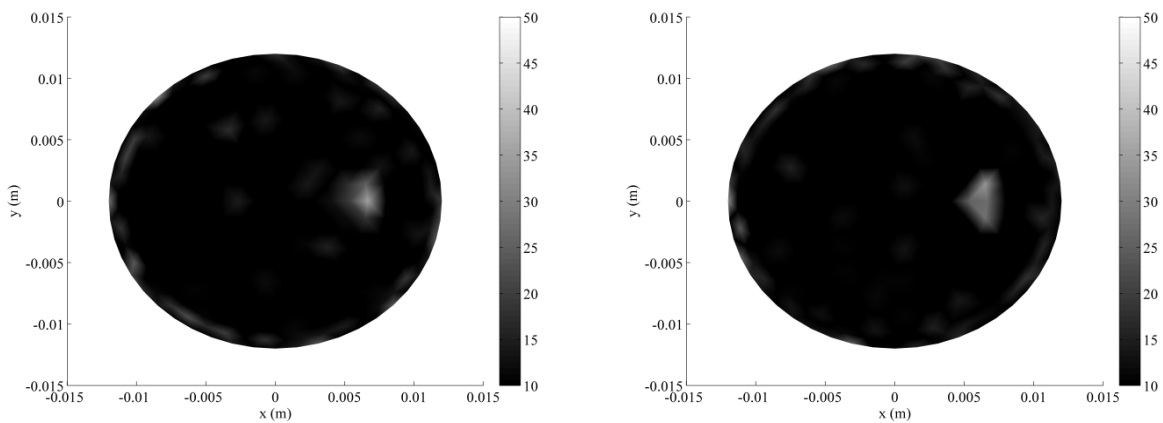

(a)            (b)

Figure 2. Reconstructed $\mu_a$ (m$^{-1}$) using (a) filtering and (b) optimization (Reference: figure 1). The reconstructions were carried out with 51 measurements and 25 MC particles in both cases.

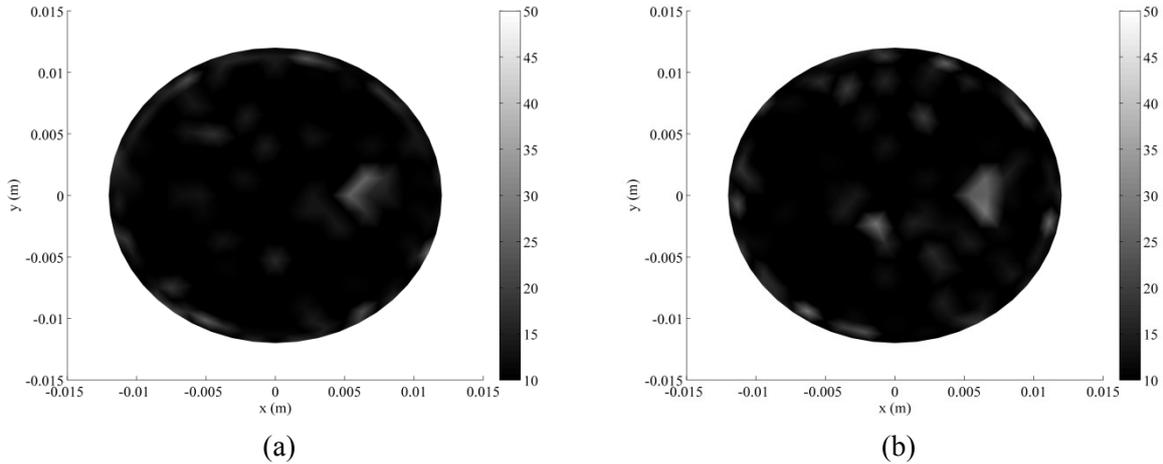

(a)            (b)

Figure 3. Reconstructed $\mu_a$ (m$^{-1}$) using (a) filtering and (b) optimization (Reference: figure 1). The reconstructions were carried out with 25 measurements and 25 MC particles in both cases.

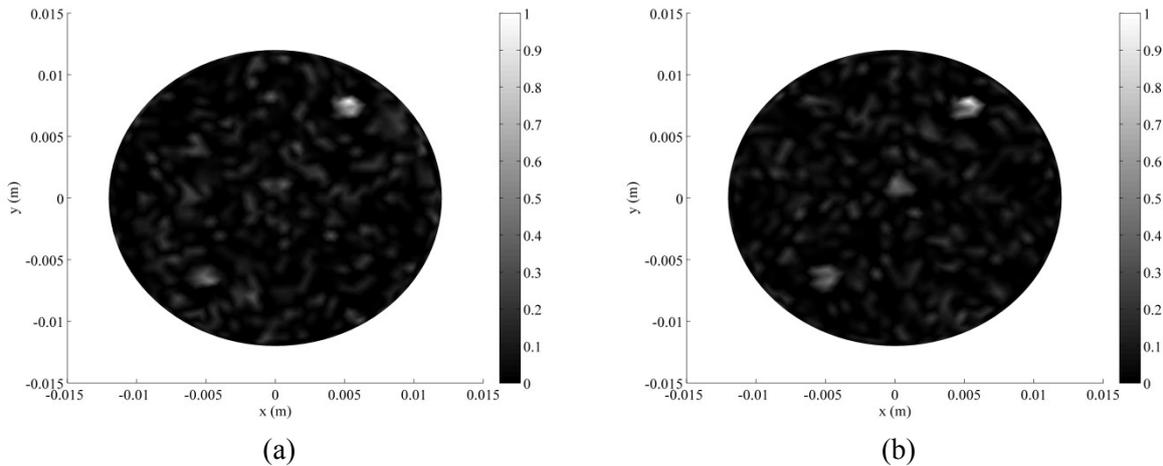

(a)            (b)

Figure 4. Relative $\mu_a$ distribution corresponding to the experimental phantom reconstructed using (a) filtering and (b) optimization

**Conclusions**

On the one hand, the presented scheme is a continuation of our recent efforts at establishing a probabilistically based evolutionary approach for global optimization with strictly rational moorings, which one may distinguish from the plentifully available mimetic approaches based on socio-cultural or biological metaphors. The proposed stochastic evolutionary scheme is based on diffusion processes and, owing to the transience of Brownian motion for states of dimension higher than 2, their evolutions are not properly bounded to remain within a prescribed domain. In future, we intend to address this issue by exploiting compensated compound Poisson or Levy processes. On the other hand, by applying the scheme to the problem of quantitative photoacoustic tomography involving both simulated and experimental measurements, we have furnished adequate numerical evidence in support of our conjecture that a global search, and not a regularized local or even a quasi-local search, is far better suited to solve realistic inverse problems with sparsely available data. The last observation may perhaps help reshape the conventional approaches that typically interpret this important class of problems as ill-posed in the sense of Hadamard and employ variously regularized quasi-Newton updates for numerical solutions.

**Data accessibility.** The MATLAB programs and the data used for the inverse problem are available in the supplementary material.

**Competing interests.** We have no competing interests.

**Authors' contributions.** The study was conceived of by DR and MV. MV and DR were also responsible for the development of the reported methodologies as well as the drafting of the manuscript. The numerical results were generated by MV. RMV was responsible for organizing the reported experiments, data collection and the final revision of the manuscript. All authors gave final approval for publication.

**Funding.** The authors have not received funding for the present work

**Acknowledgements.** We are grateful to Dr. Srirang Manohar and the biomedical photonic imaging group of the University of Twente, The Netherlands, for providing us with experimental data used to generate the results of figure 4.

## References

1. Fletcher R. 1987 *Practical Methods of Optimization.* New York, NY: John Wiley and Sons.
2. Engl HW, Hanke M, Neubauer A. 1996 Regularization of Inverse Problems. *Math. Appl.* Dordrecht, The Netherlands: Kluwer Academic Publishers Group.
3. Teresa J, Venugopal M, Roy D, Vasu RM, Kanhirodan R. 2014 Diffraction tomography from intensity measurements: an evolutionary stochastic search to invert experimental data. *J. Opt. Soc. Am. A* 31, 996-1006. (doi: 10.1364/JOSAA.31.000996)

4. Kurtz TG. 1988 Martingale problems for conditional distributions of Markov processes. *Electron. J. Probab.* **3,** 1-29. (doi: 10.1214/EJP.v3-31)

5. Stroock DW, Varadhan SRS. 1972 On the support of diffusion processes with application to the strong maximum principle. *Proc. Sixth Berkeley Symp. on Math. Statist. and Prob., Univ. of California Press.* **3,** 333-359.

6. Klebaner FC. 2001 *Introduction to Stochastic Calculus with Applications.* London, UK: Imperial College Press.

7. Sarkar S, Chowdhury SR, Venugopal M, Vasu RM, Roy D. 2014 A Kushner-Stratonovich Monte Carlo filter applied to nonlinear dynamical system identification. *Physica D: Nonlinear Phenomena* **270,** 46-59. (doi: 10.1016/j.physd.2013.12.007)

8. Kurtz TG, Ocone DL. 1988 Unique characterization of conditional distributions in nonlinear filtering. *Ann. Probab.* **16,** 80-107.

9. Eiben AE, Smith JE. 2003 *Introduction to Evolutionary Computing.* Berlin, Germany: Springer.

10. Holland JH. 1975 *Adaptation in natural and artificial systems.* Ann Arbor, MI: University of Michigan Press.

11. Kirkpatrick S, Gelatt Jr CD, Vecchi MP. 1983 Optimization by simulated annealing. *Science* **220,** 671-680. (doi: 10.1126/science.220.4598.671)

12. Kennedy J, Eberhart R. 1995 Particle swarm optimization. *Proc. IEEE Int. Conf. Neural Networks, Perth, WA* **4**, 1942-1948. (doi: 10.1109/ICNN.1995.488968)

13. Storn R, Price K. 1997 Differential evolution - a simple and efficient heuristic for global optimization over continuous spaces. *J. Global Optim.* **11,** 341-359. (doi: 10.1023/A:1008202821328)

14. Hansen N, Ostermeier A. 1996 Adapting arbitrary normal mutation distributions in evolution strategies: the covariance matrix adaptation. *Proc. IEEE Int. Conf. Evol. Comput., Nagoya* 312-317. (doi: 10.1109/ICEC.1996.542381)

15. Igel C, Hansen N, Roth S. 2007 Covariance matrix adaptation for multi-objective optimization. *Evol. Comput.* **15,** 1-28. (doi: 10.1162/evco.2007.15.1.1)

16. Sarkar S, Roy D, Vasu RM. 2014 A perturbed martingale approach to global optimization. *Phys. Lett. A* **378,** 2831-2844. (doi: 10.1016/j.physleta.2014.07.044)

17. Sarkar S, Roy D, Vasu RM. 2015 A global optimization paradigm based on change of measures. *R. Soc. open sci.* **2,** 150123. (doi: 10.1098/rsos.150123)

18. Wolpert DH, Macready WG. 1997 No free lunch theorems for optimization. *IEEE Trans. Evol. Comput.* **1,** 67-82. (doi: 10.1109/4235.585893)

19. Raveendran T, Roy D, Vasu RM. 2014 Iterated gain-based stochastic filters for dynamic system identification. *Journal of the Franklin Institute* **351,** 1093-1111. (doi:10.1016/j.jfranklin.2013.10.003)
20. Kushner HJ. 1964 On the differential equations satisfied by conditional probability densities of Markov processes, with applications. *J SIAM Control Ser. A* **2,** 106-119. (doi: 10.1137/0302009)
21. Kalman RE. 1960 A new approach to linear filtering and prediction problems. *Trans. ASME, Ser. D, Journal of Basic Engineering*, **82,** 35-45. (doi: 10.1115/1.3662552)
22. Livings DM, Dance SL, Nichols NK. 2008 Unbiased ensemble square root filters. *Physica D: Nonlinear Phenomena* **237,** 1021-1028. (doi: 10.1016/j.physd.2008.01.005)
23. Doob JL. 1957 Conditional Brownian motion and the boundary limits of harmonic functions, *Bull. Soc. Math. France* **85,** 431-458.
24. Evensen G. 2003 The ensemble Kalman filter: theoretical formulation and practical implementation. *Ocean Dynamics* **53,** 343-367. (doi: 10.1007/s10236-003-0036-9)
25. Chopin N. 2004 Central limit theorem for sequential Monte Carlo methods and its application to Bayesian inference, *Ann. Statist.* **32,** 2385-2411.
26. Osborne MJ. 2004 *An Introduction to Game Theory*. New York: Oxford University Press.
27. Nash, Jr. JF. 1950 Equilibrium points in n-person games. *Proc. Nat. Acad. Sci. USA* **36,** 48-49. (doi: 10.1073/pnas.36.1.48)
28. Revus D, Yor M. 1994 *Continuous Martingales and Brownian Motion*. Berlin: Springer.
29. Li, Q 2011 *Two Approaches to Non-Zero-Sum Stochastic Differential Games of Control and Stopping*. PhD Thesis: Columbia University.
30. Tang K, Li X, Suganthan PN, Yang Z, Weise T. 2010 *Benchmark Functions for the CEC'2010 Special Session and Competition on Large-Scale Global Optimization.*
31. Finck S, Hansen N, Ros R, Auger A. 2014 *Real-Parameter Black-Box Optimization Benchmarking 2010: Presentation of the Noiseless Functions.*
32. Wang LV, Hu S. 2012 Photoacoustic tomography: in vivo imaging from organelles to organs. *Science* **335,** 1458-1462. (doi: 10.1126/science.1216210)
33. Banerjee B, Bagchi S, Vasu RM, Roy D. 2008 Quantitative photoacoustic tomography from boundary pressure measurements: noniterative recovery of optical absorption coefficient from the reconstructed absorbed energy map. *J. Opt. Soc. Am. A* **25,** 2347-2356. (doi: 10.1364/JOSAA.25.002347)

34. Roy, D and Dash, M K, 2005 Explorations of a family of stochastic Newmark methods in engineering dynamics. Computer Methods in Applied Mechanics and Engineering 194(45), 4758-4796 (doi: 10.1016/j.cma.2004.11.010)
35. van Es P, Biswas SK, Bernelot Moens HJ, Steenbergen W, Manohar S. 2014 Initial results of finger imaging using photoacoustic computed tomography. *J. Biomed. Opt.* **19,** 060501. (doi: 10.1117/1.JBO.19.6.060501)